\newbox\grsign \setbox\grsign=\hbox{$>$} 
\newdimen\grdimen \grdimen=\ht\grsign
\newbox\laxbox \newbox\gaxbox
\setbox\gaxbox=\hbox{\raise.5ex\hbox{$>$}\llap
     {\lower.5ex\hbox{$\sim$}}}\ht1=\grdimen\dp1=0pt
\setbox\laxbox=\hbox{\raise.5ex\hbox{$<$}\llap
     {\lower.5ex\hbox{$\sim$}}}\ht2=\grdimen\dp2=0pt
\def\gax{\mathrel{\copy\gaxbox}}
\def\lax{\mathrel{\copy\laxbox}}

\documentstyle[12pt,aasms4]{article}

\received{July 1, 1996}
\accepted{August 5, 1996}

\lefthead{Mauche}
\righthead{Quasi-Coherent Oscillations}

\begin{document}

\title{The EUV Spectrum of the Quasi-Coherent\\
       Oscillations of the Dwarf Nova SS~Cygni}

\author{Christopher W.\ Mauche}
\affil{Lawrence Livermore National Laboratory,\\
       L-41, P.O.\ Box 808, Livermore, CA 94550;\\
       mauche@cygnus.llnl.gov}

\begin{abstract}
Data obtained by the {\it Extreme Ultraviolet Explorer\/} satellite are used to
determine the EUV spectrum of the quasi-coherent oscillations of the dwarf nova
SS~Cygni. It is found that the spectrum of the oscillations is neither blue nor
red nor grey relative to the net (oscillation-phase integrated) spectrum, and
hence that the oscillations cannot be explained by variations in the effective
temperature, absorbing column density, or effective area, respectively. Instead,
it is found that the amplitude of the oscillations is high at the relative
maxima of the net spectrum, and low to zero at the relative minima of the net
spectrum. This behavior can be explained by either variations in the emission
line flux atop a constant underlying continuum, or variations in the optical
depth of a haze of overlapping absorption lines, in which case the optical
depths must be $\tau\lax 1$ at the relative maxima of the net spectrum, and
$\tau\gg 1$ at the relative minima.
\end{abstract}


\clearpage   

\section{Introduction}

In a recent communication (Mauche 1996b, hereafter Paper II), we described
{\it Extreme Ultraviolet Explorer\/} ({\it EUVE\/}; \cite{bow91}; \cite{bow94}) 
observations of the 7--9~s quasi-coherent ($Q\equiv |\Delta P/\Delta t|^{-1}
> 6\times 10^4$) oscillations (``dwarf nova oscillations;'' \cite{pat81};
\cite{war5a}; \cite{war5b}) in the EUV flux of the dwarf nova SS~Cyg during an
anomalous outburst in 1993 August and a normal outburst in 1994 June/July. For
both outbursts, the period $P$ of the oscillation was observed to correlate with
the EUV intensity $I_{\rm EUV}$ according to $P\propto I_{\rm EUV}^{-0.094}$.
For a magnetospheric model to produce this variation, an effective high-order
multipole field is required with a strength at the surface of the white dwarf
of 0.1--1~MG.

Having determined the periods of the EUV oscillations of SS~Cyg from the power
spectra of the events recorded by the {\it EUVE\/} Deep Survey (DS) instrument,
we now are in a position to determine the spectrum of the oscillations from
the events recorded simultaneously by the {\it EUVE\/} Short Wavelength (SW)
spectrometer. The spectrum of the oscillations is of interest for at least
two reasons. First, one might hope to learn something about the cause of the
oscillations from the nature of their spectrum. For example, the spectrum of
the oscillations will be blue if the oscillations are due to variations in the
effective temperature; red if due to variations in the column density; grey if
simply due to variations in the effective area. At present, all that is known
about the spectrum of the oscillations is that the optical colors are bluer
than the disk ({\cite{hil81}; \cite{mid82}). Second, one might hope to learn
something about the net (oscillation-phase integrated) spectrum by studying
its temporal variations. Mauche, Raymond, \& Mattei (1995, hereafter Paper~I) 
identified many of the emission features in the net EUV spectrum of the 1993
August outburst of SS~Cyg with strong transitions of 5--7 times ionized Ne, Mg,
and Si and parameterized the continuum with a blackbody with a temperature of
$kT\approx 20$--30 eV absorbed by a neutral hydrogen column density of $N_{\rm
H}\approx 7$--$4\times 10^{19}~\rm cm^{-2}$. Failing a complete understanding
of this spectrum, any clues from its temporal variations are welcome. Is it the
case, for example, that the EUV continuum varies but that the lines do not? Such
a situation applies in U~Gem, where many of the emission lines are affected much
less than the continuum by the dips in the EUV flux at orbital phases $\phi
\approx 0.6$--0.8 (\cite{lon96}; \cite{mau6c}).

Motivated by these hopes, we determined the spectrum of the EUV oscillations
of the 1993 and 1994 outbursts of SS~Cyg. With an SW count rate of $\lax 0.3~\rm
count~s^{-1}$, an oscillation amplitude of $\sim 16\%$, and $\sim 100$ 0.5~\AA
\ independent spectral elements, we need to employ nearly every photon recorded
by the SW instrument during the $\sim 100$ kilosecond observations to construct
oscillation spectra with sufficient signal-to-noise ratio. To accomplish this
task, we need to take careful account of the varying period and phase of the
oscillation throughout the observations, and verify as best we can that other
factors (e.g., secular variations in the net spectrum) do not strongly affect
the result.

\section{Observations and Analysis}

Target-of-opportunity observations of SS~Cyg in outburst were made with {\it
EUVE\/} in 1993 August (MJD 9216.58 to 9223.12; $\rm MJD =JD - 2440000$) and
1994 June/July (MJD 9526.67 to 9529.78 and 9532.54 to 9536.94). The optical and
DS count rate light curves of the outbursts are shown in Figure~1 of Mauche
(1996a). On both occasions, the optical flux was above $V=10$ for $\approx 16$
days, but the 1993 outburst was anomalous in that it took $\approx 5$ days for
the light curve to reach maximum, whereas typical outbursts (such as the 1994
outburst) reach maximum in 1 to 2 days.

As described in Paper~II, we determined the period of the oscillation of the EUV
flux of SS~Cyg during each valid interval by calculating the power spectrum of
DS count rate light curves with 1~s time resolution. The period $P$ of the
oscillation as a function of time and of the log of the 75--120~\AA \ SW count
rate is shown in Figures~1 and 2 of Paper~II. The phase $\phi _0$ and amplitude
of the oscillation during each valid time interval was calculated by
phase-folding the DS data on the period appropriate to that interval and fitting
a function of the form $f(\phi )=A+B\sin 2\pi (\phi + \phi_0)$ where $\phi =
t/P$. The relative amplitudes $B/A$ of the oscillation are shown as a function
of the log of the 75--120~\AA \ SW count rate in Figure~1. The weighted mean of
these amplitudes is $16.1\%\pm 0.3\%$ ($\chi^2/{\rm dof}=129/60$) for the 1993
outburst, and $14.7\%\pm 0.2\%$ ($\chi^2/{\rm dof}=609/72$) for the 1994
outburst. Note, however, that the amplitudes during the 1994 outburst are often
low when the SW count rates are above $\sim 0.6~\rm counts~s^{-1}$; in a plot
of the amplitude as a function of the period, the amplitudes are
{\it systematically\/} low when the period is below $\sim 7.5$~s: above 7.5~s,
the weighted mean of the amplitudes is $17.7\% \pm 0.2\%$ ($\chi^2/{\rm dof}
=194/58$); below 7.5~s, it is $10.8\%\pm 0.3\%$ ($\chi^2/{\rm dof} =54.0/13$).
This magic period of $\sim 7.5$~s is (perhaps not coincidentally) the same as
the minimum asympototic period reached during the plateau of the 1993 outburst.
Apparently, the oscillation can be driven below this magic period, but only
for a short time (see Fig.~1 of of Paper~II) and only at the expense of its
amplitude. Below this magic period, the amplitude of the oscillation is
reasonably constant with count rate at $\sim 16\%$, but the large reduced
$\chi ^2$s demonstrate the orbit-to-orbit variability about this value.

To investigate possible spectral variability during the observations, we divided
the 75--120~\AA \ SW bandpass at 90~\AA , determined the count rate during each
valid interval in the 75--90~\AA \ (``hard'') and 90--120~\AA \ (``soft'')
channels, and calculated the ratio of the hard over soft count rates; these
count rate ratios are shown as a function of the total 75--120~\AA \ count rate
in Figure~2. The weighted mean of these ratios is $0.654\pm 0.007$ ($\chi^2/{\rm
dof}= 70.8/60$) for the 1993 outburst and $0.629\pm 0.006$ ($\chi^2/{\rm
dof}=105/72$) for the 1994 outburst, and the difference between these values is
$0.025\pm 0.009$. These results demonstrate that the ratio of the hard/soft
count rates are reasonably constant with time and count rate, that the spectrum
--- unlike the amplitude --- is not affected when the period of the
oscillation is below $\sim 7.5$~s, and that the ratio of the hard over soft
count rates of the 1993 and 1994 outbursts differs by $<3\sigma $. Evidently,
the EUV spectrum of SS~Cyg is independent of total intensity, oscillation period,
oscillation amplitude, or even outburst type; the spectrum might as well be cut
out of titanium for all it changes.

Encouraged by the fact that the spectrum of the EUV oscillations of SS~Cyg is
reasonably constant with time, we used the periods and phases determined from
the DS data to accumulate SW spectra during the high and low portions of the
oscillation. Specifically, photons satisfying $\sin 2\pi (\phi + \phi_0) > 0$
where accumulated in one array, and those satisfying $\sin 2\pi (\phi + \phi_0)
\le 0$ where accumulated in another. We also tried accumulating only those
photons from near the peak and valley of the oscillation, specifically those
photons satisfying $\sin 2\pi (\phi + \phi_0)>+0.5$ and $<-0.5$, but this
procedure wasted nearly a third of the photons, and produced a result which was
the same within the errors. The spectrum of the high and low portions of the
oscillation of the 1993 (1994) outburst of SS~Cyg is shown in the top panel of
Figure~3 (4). The spectrum of the oscillations is the difference between these
two spectra shown as filled squares with error bars in the middle panel of the
figure. For comparison, we also show the sum (net; oscillation-phase integrated)
spectrum after scaling by a factor of $s=0.1064$ (0.1059). This scale factor
corresponds to an oscillation amplitude of $B/A=\pi s/2=16.7\%$ (16.6\%), which
reasonably well reproduces the count-weighted mean oscillation amplitude
measured by the DS instrument of 16.8\% (16.2\%). The ratio of the high-phase
spectrum to the low-phase spectrum is shown in the bottom panel of the figure
as residuals about the weighted mean ratio of 1.218 (1.199). With $\chi^2/{\rm
dof}$ of 139/115 (182/115), the hypothesis that the high-phase spectrum
differs from the low-phase spectrum simply by a scale factor can be rejected
with $\gax 90\%$ ($\gax 99.9\%$) confidence. Simply stated, the EUV oscillations
of SS~Cyg are not grey relative to the oscillation-phase integrated spectrum.

While the EUV oscillations of SS~Cyg are not grey, they also clearly are not
red or blue: the residuals shown in the bottom panels of Figures 3 and 4 do not
vary systematically from one end of the spectrum to the other. This statement
can be quantified for two simple models. Consider variations in the absorbing
column density: as the variation in $N_{\rm H}$ increases from zero, the
amplitude of the oscillation increases, but the ratio spectrum becomes
increasingly red. Assuming that $N_{\rm H}(\phi )=N_{\rm H_0}-N_{\rm H_1} \sin
2\pi\phi $, $\chi^2$ is minimized for $N_{\rm H_1}=4\times 10^{18}~\rm cm^{-2}$,
but $\chi^2/{\rm dof} = 171/115$ (183/115), so the hypothesis that the EUV
oscillations are produced by variations in the absorbing column density can be
rejected with $\gax 99.9\%$ confidence. Next, consider variations in the
effective temperature: as the variation in $kT$ increases from zero, the
amplitude of the oscillation again increases, but the ratio spectrum becomes
increasingly blue. Assuming that $kT(\phi)=kT_0+kT_1 \sin 2\pi\phi $ and
$kT_0=20$~eV, $\chi^2$ is minimized for $kT_1=0.5$~eV (0.4~eV), but $\chi^2/{\rm
dof}=149/115$ (202/115); for $kT_0=30$~eV, $\chi^2$ is minimized for
$kT_1=1.0$~eV (0.8~eV), but $\chi^2/ {\rm dof}=145/115$ (204/115); in both
instances, the hypothesis that the EUV oscillations are produced by variations
in the effective temperature can be rejected with $\gax 95\%$ ($\gax 99.9\%$)
confidence. 

\section{Discussion}

Having shown formally that the spectrum of the EUV oscillations of SS~Cyg is
neither blue nor red nor grey relative to the oscillation-phase integrated
spectrum, it is useful to refer to Figures~3 and 4 to get a qualitative sense
of the wavelength dependence of the oscillations. First, consider those bins
in the difference spectra which are $\le 0$ within the errors: with only a few
exceptions, they invariably lie at relative minima in the sum spectra: e.g., at
$\approx 74$, 82, 87, 96, 105~\AA . In contrast, those bins in the difference
spectra which exceed the (scaled) sum spectra invariably lie at the relative
maxima in the sum spectra: e.g., at $\approx 86$, 92, 100~\AA . Overall, it is
clear that the flux at the relative minima in the sum spectra is constant
with oscillation phase, and that it is the flux at the relative maxima which
oscillates. 

How does this result constrain the cause of the oscillations? Unfortunately,
the answer is as uncertain as the processes responsible for the shape of the
oscillation-phase integrated spectrum. In the simplest interpretation, the net
spectrum is composed of emission lines superposed on a weak continuum. In that
case, the lines oscillate and the continuum does not. This interpretation is
favored by our  success in identifying numerous emission lines in the net
spectrum with strong transitions of 5--7 times ionized Ne, Mg, and Si (Paper~I).
On the other hand, we cannot yet rule out the possibility that the net spectrum
is formed by complex radiation transfer effects in the outflowing wind known
from {\it IUE\/} observations to exist when SS~Cyg is in outburst (\cite{hea78}).
In that case, the supposed emission lines in the net spectrum are instead
regions of relative transparency in a haze of overlapping absorption lines.
Then, variations in the optical depth of the haze will most strongly affect
those regions of the spectrum where $\tau\lax 1$ --- the relative maxima --- and
least strongly affect those regions where $\tau\gg 1$ --- the relative minima.
Since the spectrum of the EUV oscillations does not distinguish between these
alternatives, we defer further analysis of the EUV spectrum of SS~Cyg to a later
work.

\acknowledgments

The {\it EUVE\/} observations of SS~Cyg could not have been accomplished
without the efforts of the members, staff, and director, J.\ Mattei, of the
American Association of Variable Star Observers; {\it EUVE\/} Deputy Project
Scientist Ron Oliversen; and the staffs of the Center for EUV Astrophysics
(CEA), the {\it EUVE\/} Science Operations Center at CEA, and the Flight
Operations Team at Goddard Space Flight Center. This work was performed under
the auspices of the U.S.\ Department of Energy by Lawrence Livermore National
Laboratory under contract No.~W-7405-Eng-48.

\clearpage    


\clearpage    


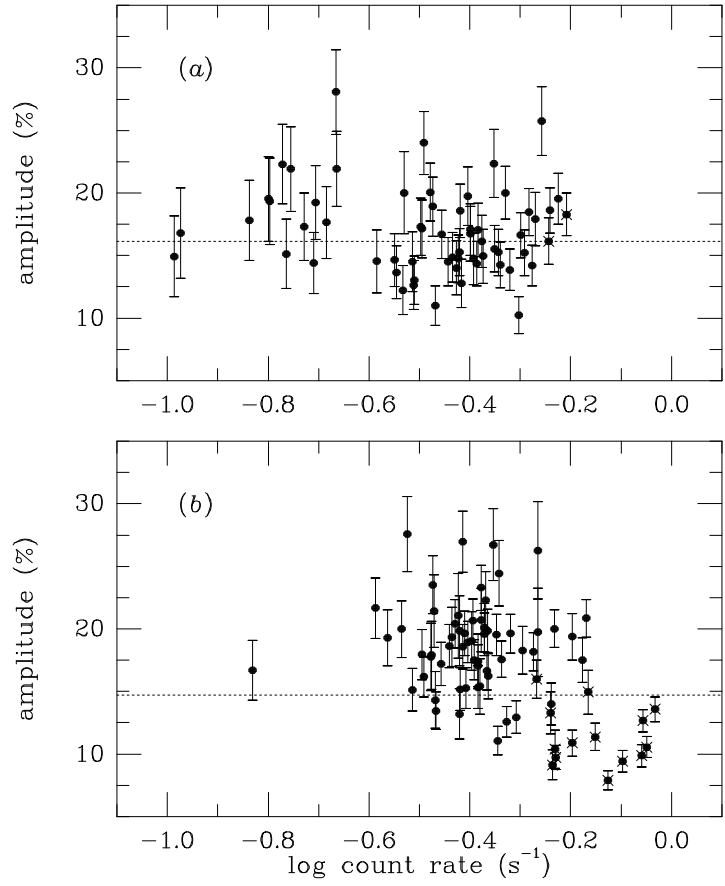
\begin{figure}
\caption{Amplitude of the EUV oscillations as a function of the log of the
75--120 \AA \ SW count rate for the ({\it a\/}) 1993 August and ({\it b\/})
1994 June/July outbursts of SS~Cyg. Data points for which $P\le 7.5$~s are
shown with crosses.}
\end{figure}

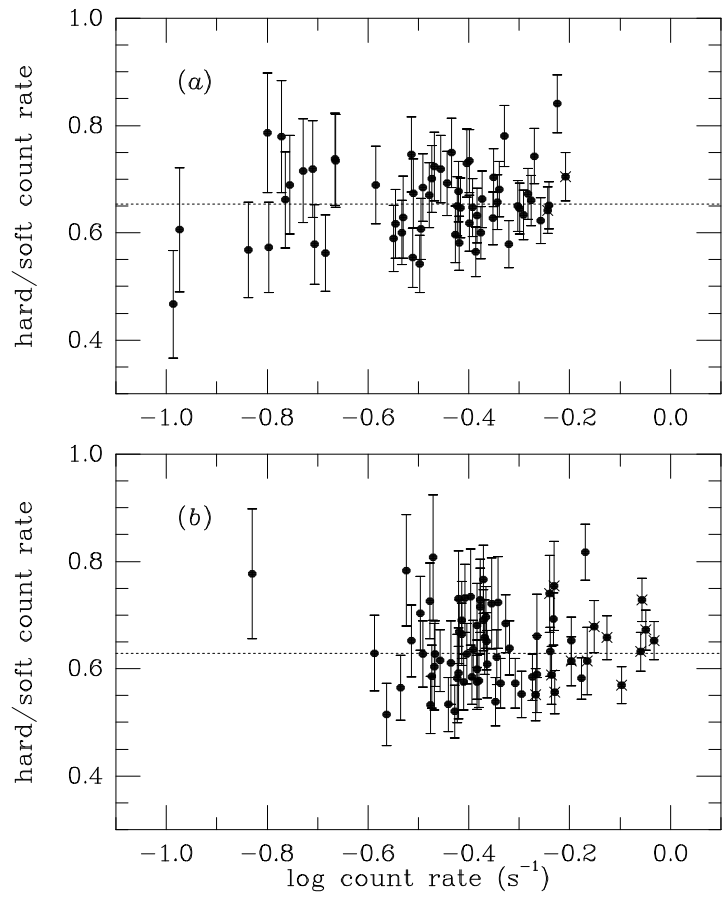
\begin{figure}
\caption{Ratio of the hard (75--90~\AA) over soft (90--120~\AA ) count rates
as a function of the log of the 75--120 \AA \ count rate for the ({\it a\/})
1993 August and ({\it b\/}) 1994 June/July outbursts of SS~Cyg. Data points
for which $P\le 7.5$~s are shown with crosses.}
\end{figure}

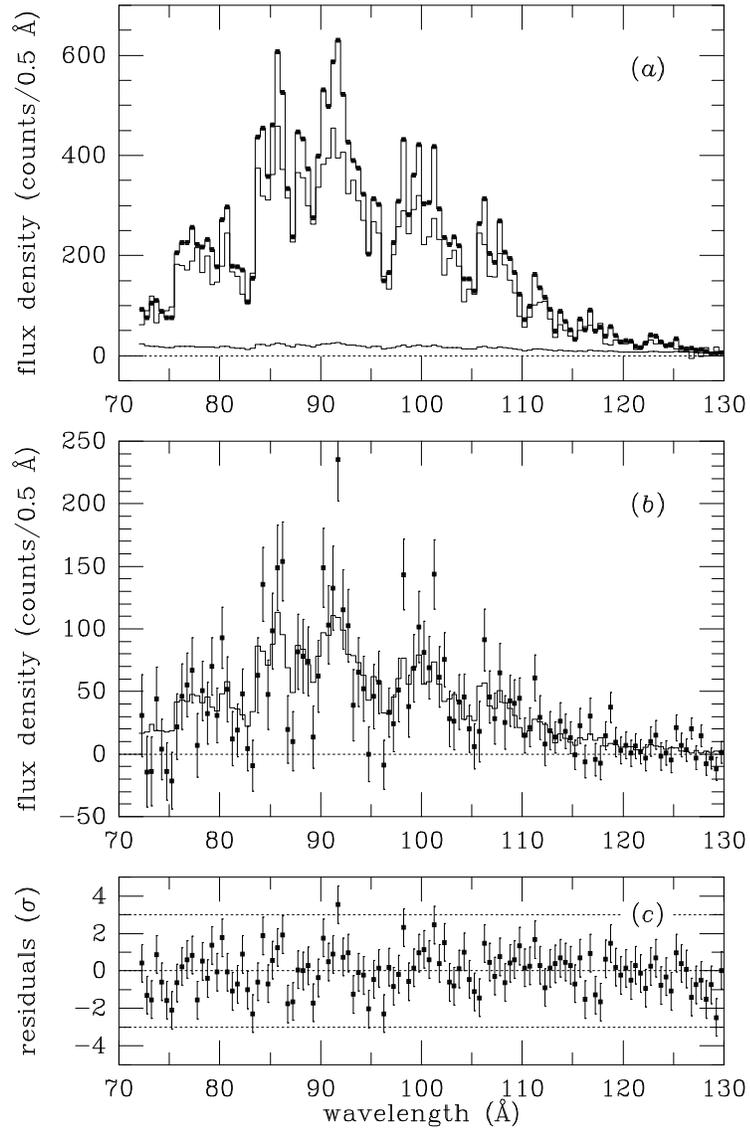
\begin{figure}
\caption{({\it a\/}) SW count spectrum for the 1993 August outburst of SS~Cyg.
Upper histogram with solid square points is the spectrum accumulated during
the high phase of the oscillation; lower histogram is the spectrum accumulated
during the low phase of the oscillation. Trace at the bottom of the figure
plots the $1\sigma $ error vector associated with the high-phase spectrum.
({\it b\/}) Difference spectrum ({\it solid squares with error bars\/}) and
sum spectrum scaled by 0.1064 ({\it histogram\/}). ({\it c\/}) Ratio of the
high-phase spectrum over the low-phase spectrum plotted as residuals about
1.218.}
\end{figure}

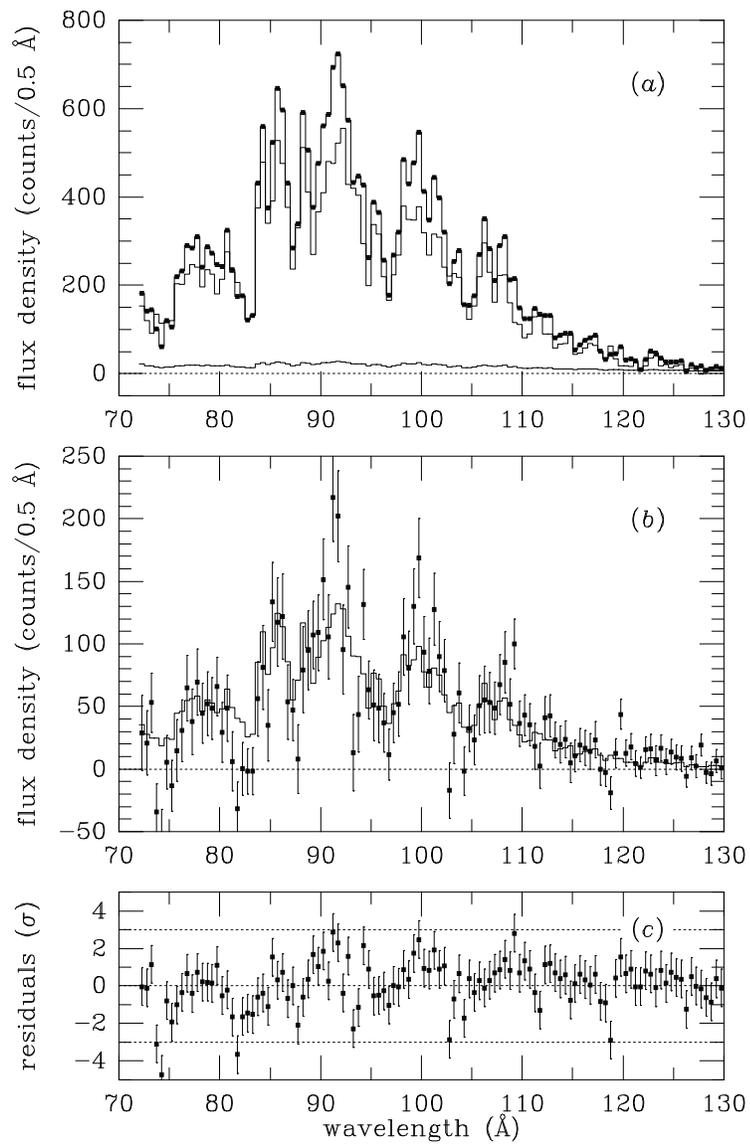
\begin{figure}
\caption{Similar to Fig.~3, for the 1994 June/July outburst of SS~Cyg. The
scale factor for the sum spectrum is 0.1059 and the residuals are plotted about
1.199.}
\end{figure}

\end{document}